\documentclass[onecolumn]{revtex4}
 \usepackage{verbatim}
\usepackage{graphicx}% Include figure file
\usepackage{float}
\usepackage[utf8]{inputenc}
\usepackage[french,english]{babel}
\usepackage{amssymb}
\usepackage{amsfonts}
\usepackage{hyperref}
\usepackage{color}

\newcommand{\be}{\begin{eqnarray}}
\newcommand{\ee}{\end{eqnarray}}
\newcommand{\dif}{\mathrm{d}}

\newcommand{\C}{\mathbb{C}}

\newcommand{\Z}{\mathbb{Z}}

 \newcommand{\im}{\mathbf{i}}
  
 \newcommand{\abs}[1]{\left\vert#1\right\vert}

\begin{document}
\title{Extreme value statistics of 2d Gaussian Free Field: effect of finite domains}
\author{Xiangyu Cao}
\affiliation{LPTMS, CNRS (UMR 8626), Université Paris-Saclay, 91405 Orsay, France}
\thanks{Corresponding author}
\email{xiangyu.cao08@gmail.com}
\author{Alberto Rosso} 
\affiliation{LPTMS, CNRS (UMR 8626), Université Paris-Saclay, 91405 Orsay, France}
\email{alberto.rosso@u-psud.fr}
\author{Raoul Santachiara}
\affiliation{LPTMS, CNRS (UMR 8626), Université Paris-Saclay, 91405 Orsay, France}
\email{raoul.santachiara@u-psud.fr}
\begin{abstract}
We study minima statistics of the 2d Gaussian Free Field on circles in the unit disk with Dirichlet boundary condition. Free energy distributions of the associated Random Energy models are exactly calculated in the high temperature phase, and shown to satisfy the duality property, which enables us to predict the minima distribution by assuming the freezing scenario. Numerical tests are provided. Related questions concerning the GFF on a sphere are also considered.
\end{abstract}
\maketitle

\section{Introduction}\label{intro} 
%The 2D Gaussian Free Field (GFF) is fundamental 
The minima statistics of the 2D Gaussian Free Field (GFF) is a fundamental problem relevant in different contexts, ranging from multi-fractal behaviours in quantum disordered systems \cite{chamon1996localization,castillo97dirac,fyodorov2015high}, rare events in climate science \cite{hubert1993multifractals} to zeros of Riemann zeta function \cite{arguin2015maxima}. While rigorous results exist for the leading behaviour of extrema and near extrema of GFF, \textit{c.f.} \cite{ding2014extreme} and references therein, less is known about the full extremum distribution. This was calculated only in two cases: the GFF on a circle by Fyodorov and Bouchaud \cite{fyodorov08rem} (FB), and on an interval \cite{fyodorov2009statistical}. In both cases, the authors studied the associated Random Energy Model \cite{derrida1980random,derrida1981random} with logarithmically correlated energy landscape. They showed that the minima distribution can be obtained by a \textit{freezing scenario} describing the low temperature phase of those Random Energy Models \cite{carpentier2001glass}. Moreover the authors of \cite{fyodorov2009statistical} pointed out the key rôle of a \textit{duality property}, reminiscent of well-known dualities in Conformal Field Theory \cite{zamolodchikov1996conformal,duplantier06dual}. 

In general, the GFF $\phi_\Sigma(x)$ on a 2D domain $\Sigma$ is determined by its \textit{Green function} $\overline{ \phi_\Sigma(x)\phi_\Sigma(y)} = G_\Sigma(x,y) \approx -2\ln|x-y|$. It has both a short distance (ultraviolet) and a long distance (infra red) divergence. While the first is at the origin of the freezing scenario, the second must be cured in some way. For instance, one may take $\Sigma$ to be the unit disk and specify the Dirichlet boundary condition. From this point of view, the results in \cite{fyodorov08rem} should be interpreted as the limit of vanishing circle radius. In this work, we study GFF on circles of any radius, where the minimum distribution is expected to be affected by the presence of boundary, \textit{c.f.} \cite{fyodorov2009statistical}, sect. 6.2.1. Alternatively, $\Sigma$ can be a compact surface, such as a sphere. One of our main motivations is to see whether freezing and duality are robust in these different situations.  

When $\Sigma$ is a compact surface, the Green function $G_\Sigma$ is the solution kernel to the Poisson equation. For the sphere $\Sigma = \mathbb{S}$, with the standard metric, we have 
 (see \textit{e.g.}, \cite{david2014liouville}) 
\begin{equation}\label{David-G}
G_{\mathbb{S}}(z,w) =  - \ln \frac{|z - w|^2}{(1+|z|^2)(1+|w|^2)} - 1
\end{equation}
where we identify $\mathbb{S}$ with $\C\cup\{\infty\}$ by stereographic projection.
On a latitude $\{|z| = r\}$ and noting $\phi_0(\theta):= \phi_\mathbb{S}(r e^{\im\theta})$, 
\begin{equation}\label{G-cir-sph}
\overline{\phi_0(\theta) \phi_0(\theta')} = -2\ln| e^{\im\theta} -  e^{\im\theta'}| + a.
\end{equation}
Here $a$ is a \textit{zero mode}, which contributes a convolution with an independent Gaussian in terms of minima distribution. We shall set $a$ to zero. Then the minima distribution of the GFF on a latitude on $\mathbb{S}$ does not depend on $r$, but always coincides with the FB case.

As we will see, this is no longer true for the unit disk $\mathbb{D}$. the Green function on $\mathbb{D}$ with Dirichlet boundary condition is \cite{duplantier09kpz}%-2 \ln\tanh\frac{d(z,w)}{2} =
$G_{\mathbb{D}}(z,w)  =  -2\ln\abs{\frac{z-w}{1 - z\overline{w}}}.$
On a centred circle of radius $ \sqrt{q}$, $q\in (0,1)$, noting $\phi(\theta):= \phi_{\mathbb{D}}(\sqrt{q} e^{\im\theta})$
\begin{equation}\label{G_cir_eq}  
\overline{\phi(\theta) \phi(\theta')}= -2 \ln\abs{\frac{1 - z}{1-qz}} + a', \;  z := e^{\im (\theta-\theta')}, 
\end{equation}
where $a'$ is the zero mode that we will set to zero as before. In terms of Fourier modes, 
\begin{equation}
\phi(\theta) = \Re \sum_{k \neq 0} \sqrt{\mu_k} \exp(\im\theta k) \mathsf{N}_k,  \; 
 \mu_k := \frac{1}{|k|}(1 - q^{|k|}), \;  k\neq0 \label{eq-modes}
\end{equation}
Here $\mathsf{N}_k$ are complex numbers whose real and imaginary parts are i.i.d standard normal random variables, and $\Re z$ is the real part of $z$. When $q\rightarrow 0$ the model eq. (\ref{G_cir_eq}) reduces to the FB case $\phi_0$. 
Note that the model defined by eq. (\ref{eq-modes}) makes perfect sense also for $q\in [-1,0)$. For this range of $q$, eq. (\ref{G_cir_eq}) can be obtained from the Green function $G_{\mathbb{P}}(z,w) = G_{\mathbb{S}}(z,w) - G_{\mathbb{S}}(z,\hat{w})$ (where $\hat{w} = -\overline{w^{-1}}$ is the antipodal of $w$), by taking $|z| = |w| = \sqrt{-q}$. We can then interpret the results for $q < 0$ as the minimum of $\phi_{\mathbb{P}}(z) := (\phi_\mathbb{S}(z) -\phi_\mathbb{S}(\hat{z}))/\sqrt{2}$ on a latitude of $\mathbb{S}$. In particular, for $q = -1$, we are considering the extrema of $(\phi_0(\theta) - \phi_0(\theta + \pi))/\sqrt{2}$, $\phi_0$ being that of the FB model.

In the sect. \ref{calc_sec}, we will solve the model defined by eq. (\ref{G_cir_eq}); sect. \ref{sect-num} is devoted to numerical tests of the solution.   

\section{Distribution of Minima}\label{calc_sec}
Let us begin by clarify the term ``minima of GFF''. Indeed, one defines first the minima of \textit{finite} systems $y_M := \min_{i=1}^M \phi_i$. Here $\phi_i = \phi\left(\frac{2\pi i}{M}\right)$ and $\phi$ is defined by eq. (\ref{eq-modes}) where the sum is cut to $|k| \leq \frac{M}{2}$. We are interested in the rescaled minima $y = \lim_{M\rightarrow\infty} \frac{y_M - a_M}{b_M}$, where the scaling behaviours $a_M = -2\ln M + \frac{3}{2}\ln\ln M + O(1)$ and $b_M = 1 + O\left(\frac{1}{\ln M}\right)$ were predicted in \cite{carpentier2001glass}.  

To calculate the distribution of $y$, we study the Random Energy Model defined by the partition function $\mathcal{Z}_M = \sum_{i=1}^M e^{-\beta\phi_i}$. $\beta=T^{-1}$ is the inverse temperature. Its positive moments (or replica averages) $\overline{\mathcal{Z}_M^n}$ ($n = 2, 3 \dots$) can be calculated by applying Wick theorem and using eq. (\ref{G-cir-sph}). For each $n$, for $\beta$ small enough, the sum can be approximated by a Coulomb gas integral as $M\rightarrow \infty$:
\begin{eqnarray}
\overline{\mathcal{Z}_M^n} \doteq \left( M e^{\frac{\beta^2}{2} \overline{\phi_i^2}} \right)^n  Z_n, \; \overline{\phi_i^2} = \sum_{|k| \leq \frac{M}{2}} \mu_k  \label{ZMn_eq} \end{eqnarray}
where 
\begin{eqnarray}
Z_n := \int_{0}^{2\pi} \prod_{i = 1}^n \frac{\dif \theta_i}{2\pi} 
\prod_{i<j} \abs{\frac{z_i - z_j}{1 - q z_i \overline{z_j}}}^{-2\gamma},\;
z_i := e^{\im \theta_i},\;  \gamma := \beta^2. \label{Zn_eq}
\end{eqnarray}
This integral converges for $\beta < \beta_{c,n} =  n^{-1/2}$, and this is precisely the condition for eq. (\ref{ZMn_eq}) to hold, \textit{c.f.} \cite{fyodorov08rem} sect 3.1. The temperature $\beta_{c,n}$'s, called pre-freezing temperatures, are precursors of a true transition separating a high-$T$ phase and a glassy phase. To see this heuristically, one can compare the number of energy levels, $M$, and the mean Boltzmann weight of each level, $\propto M^{\beta^2}$ \footnote{Indeed, $\overline{e^{-\beta\phi_i}} = \exp\left(\frac{\beta^2}{2} \overline{\phi_i^2}\right)$, but $\overline{\phi_i^2} =  2\ln M + O(1)$, \textit{c.f.} eq. (\ref{eq-modes}).} A transition occurs at $\beta_c = \beta_{c,1} = 1$:
\begin{itemize}
\item[-] When $\beta<1$, $M \gg M^{\beta^2}$, $\mathcal{Z}_M$ is entropy dominated. So we expect that free energy distribution in the thermodynamic limit can be obtained by analytically continuing the Coulomb-gas integral (\ref{Zn_eq}), which we do in sect. \ref{sect-moments}. 
\item[-] When $\beta > 1$,  $M^{\beta^2} \gg M$, $\mathcal{Z}_M$ is dominated by a few ($\sim O(1)$) low energy levels and the leading term ($\propto \ln M$) of the entropy vanishes. This non-analyticity extends to the full free energy distribution, and invalidates any analytic expansion method. The freezing scenario allows to extend non-analytically $\beta < 1$ results to $\beta > 1$, and to predict the distribution of $y$ at $\beta \rightarrow \infty$, see sect. \ref{sec-freez}.
\end{itemize}

\subsection{Analytic continuation of moments (when $\beta<1$)}\label{sect-moments}
The analytical expansion $Z_n$ will be done using Jack polynomials \cite{jack1970class} (following conventions of \cite{Lesage1995csm}). Denoted by $J_\lambda(z;\alpha)$, they depend on variables $z=(z_1, \dots, z_n)$, a parameter $\alpha$ and a \textit{partition} $\lambda$. The latter is essentially a \textit{Young diagram}, \textit{i.e.}, a set of unit square boxes with integer coordinates, \textit{c.f.} fig (\ref{fig-russian}
\be \lambda \equiv \{ (x,y): 0 \leq x < \lambda_y,  0 \leq y < l(\lambda) \} \ee
where $\lambda_0 \geq \lambda_1 \geq \dots\geq\lambda_{l(\lambda)-1} > 0$ ($l(\lambda)$ is its \textit{length}). We denote by $|\lambda| := \sum_{j = 0}^{l(\lambda)-1} \lambda_j$ the total number of boxes in $\lambda$, \textit{i.e.}, its size.

An important property of the Jack polynomials is the Cauchy identity \cite{stanley1989jack}
\begin{equation}
\prod_{i,j = 1}^n (1 - q z_i w_j)^{-\alpha} = \sum_\lambda q^{|\lambda|} J_{\lambda}(z;\alpha)
J_{\lambda}(w;\alpha) j^{-1}_\lambda(\alpha), \label{eq_cauchy0}
\end{equation}
where the RHS sums over all the partitions. The value of $j_\lambda(\alpha)$ will turn out irrelevant. Setting $\alpha = -\gamma$, $z_i =  e^{\im \theta_i}$ and $w_i =  e^{-\im \theta_i}$, we can use eq. (\ref{eq_cauchy0}) to rewrite the denominator of integrand of eq. (\ref{Zn_eq}) as
\begin{equation}\label{eq_Zn_denom}
\prod_{i<j}\abs{\frac{1}{1 - q z_i \overline{z_j}}}^{-2\gamma} = (1 - q)^{-n\gamma}
\sum_\lambda  \frac{J_{\lambda}(z;\alpha)
J_{\lambda}(\overline{z};\alpha)}{ j_\lambda(\alpha) }q^{|\lambda|}
\end{equation}
where $\overline{z} = (\overline{z_1}, \dots, \overline{z_n})$. We shall then need the inner product norm \cite{Macdonald88anew}
\begin{eqnarray}
\int_{0}^{2\pi}
\prod_{i = 1}^n \frac{\dif \theta_i}{2\pi} 
\abs{\Delta}^{2\alpha} 
 \frac{J_{\lambda}(z;\alpha)
J_{\lambda}(\overline{z};\alpha)}{c_n(\alpha) j_\lambda(\alpha) }
=  \prod_{(x,y)\in\lambda}g_n(x,y;\alpha), \label{eq_norm} \\
 c_n(\alpha) = \int_{0}^{2\pi} \prod_{i = 1}^n \frac{\dif \theta_i}{2\pi} \abs{\Delta}^{2\alpha}
  = \frac{\Gamma(1 + n\alpha)}{\Gamma(1 + \alpha)^n} 
 \label{eq_cn} \\  
 g_n(x,y;\alpha) = \frac{x + \alpha(n - y)}{x + 1 + \alpha(n - y - 1)}. \label{eq_gn}
\end{eqnarray}
where $\Delta := \prod_{i<j}(z_i - z_j)$, and $c_n$ is the Dyson integral \cite{dyson1962statistical1}. Combining (\ref{eq_Zn_denom}) and (\ref{eq_norm}), we obtain 
\begin{equation}
Z_n(\beta, q) = (1 - q)^{-n\gamma}  c_n(-\gamma)  \sum_\lambda q^{|\lambda|}\prod_{(x,y)\in\lambda}  
g_n(x,y;-\gamma) \label{eq_Zn1}
\end{equation} 
This equation makes sense for $n$ and $\beta$ generic, which allows us to extend the definition $Z_n(\beta, q)$ beyond the domain where corresponding Coulomb-gas integrals exist.

Now we define the fluctuation part $f$ of the free energy $\mathcal{F}_M = -\beta^{-1} \ln \mathcal{Z}_M$ as
\begin{equation}\label{eq_def_f}
f := -\beta^{-1}\ln z, \; z :=  \frac{\mathcal{Z}_M}{Z_e}, \; Z_e := M e^{\frac{\beta^2}{2} \overline{\phi_i^2}} \frac{(1 - q)^{-\gamma}}{\Gamma(1-\gamma)}
\end{equation}
So $f$ and $\mathcal{F}_M$ differ by a shift, containing the $M$-dependent part, so that the distribution of $f$ has a well-defined limit as $M \rightarrow  \infty$. Indeed, writing $t = -n \beta$, one verifies that the equations (\ref{ZMn_eq}), (\ref{eq_Zn1}) and (\ref{eq_def_f}) imply (at $M = \infty$)
\begin{eqnarray}
\overline{\exp{t f}} = \Gamma(1 + t\beta) \mathbf{s}(t,\beta,q),\label{eq_etf} \\
\mathbf{s}(t,\beta,q) := \sum_\lambda q^{|\lambda|} 
\prod_{(x,y)\in\lambda} \frac{x\beta + y\beta^{-1} + t}{(x+1)\beta + (y+1)\beta^{-1} + t }  \label{eq_s}.
\end{eqnarray}
Note that when $q = 0$, $\mathbf{s} \equiv 1$ and $\overline{\exp{t f}} =  \Gamma(1 + t\beta)$, coinciding with the FB solution. The series $\mathbf{s}(t,\beta,q)$ encodes the effects of the finite domain, and is the main result of this work (supplemented by an efficiently calculable \textit{matrix product} rewriting, \textit{c.f.} \ref{mp-append}). Our approach relying on formal analytical continuation of eq. (\ref{eq_Zn1}) and (\ref{ZMn_eq}) is non-rigorous and the result eq. (\ref{eq_etf}) is a conjecture. Its validity is well supported by careful numerical tests (sect. \ref{sect-num}). Moreover, in \ref{sec_expan} we check that it reproduces correctly a non-trivial coefficient in the high-$T$ expansion. To prove rigorously eq. (\ref{eq_etf}) we should compute the high-$T$ expansion to all order, similarly as in \cite{ostrovsky2008intermittency,ostrofsky09mellin}, but this project is left to the future.
 
\subsection{Freezing and duality (when $\beta>1$)}\label{sec-freez}
\begin{figure}
\center
\includegraphics[scale=.3]{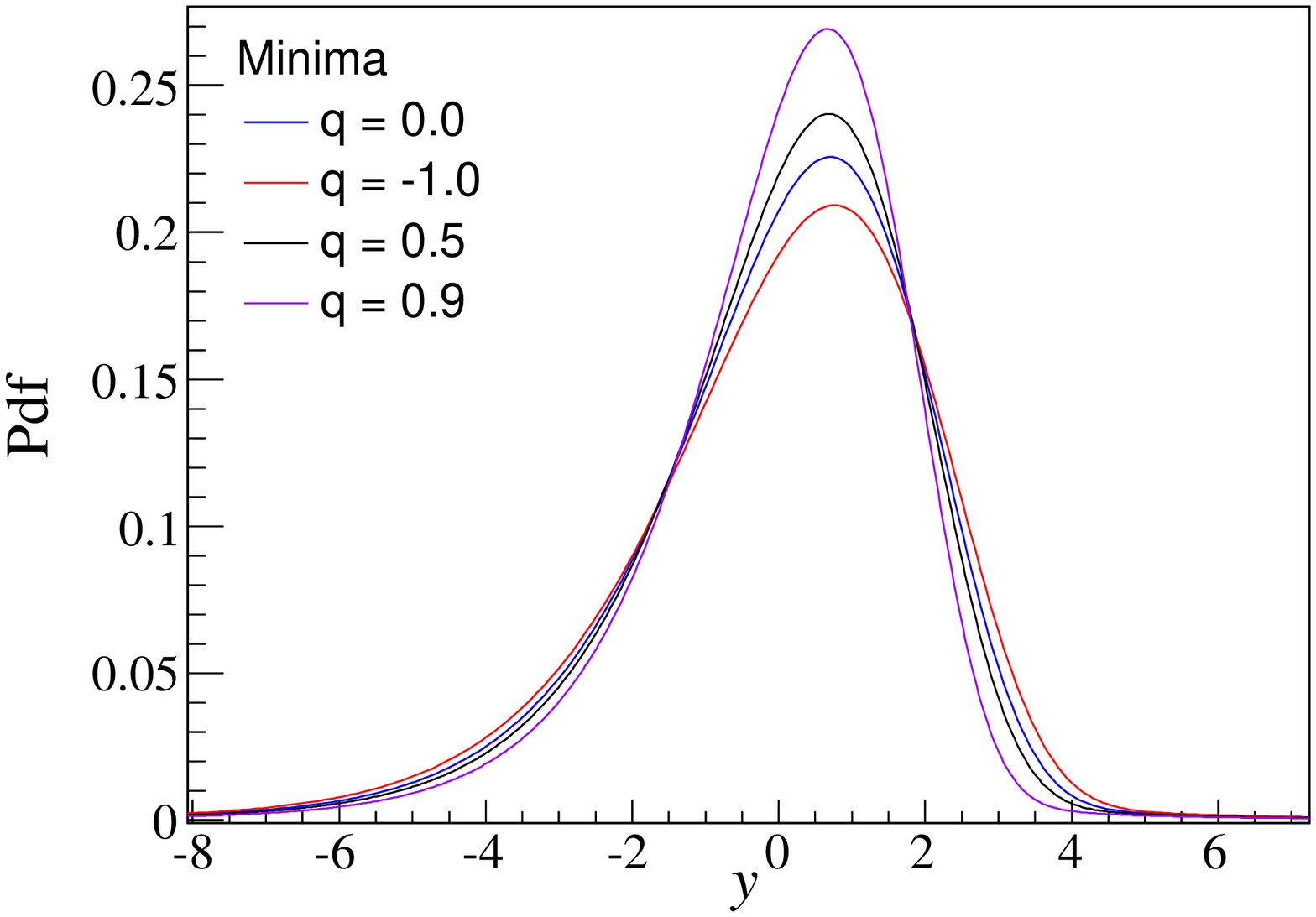}
\includegraphics[scale=.3]{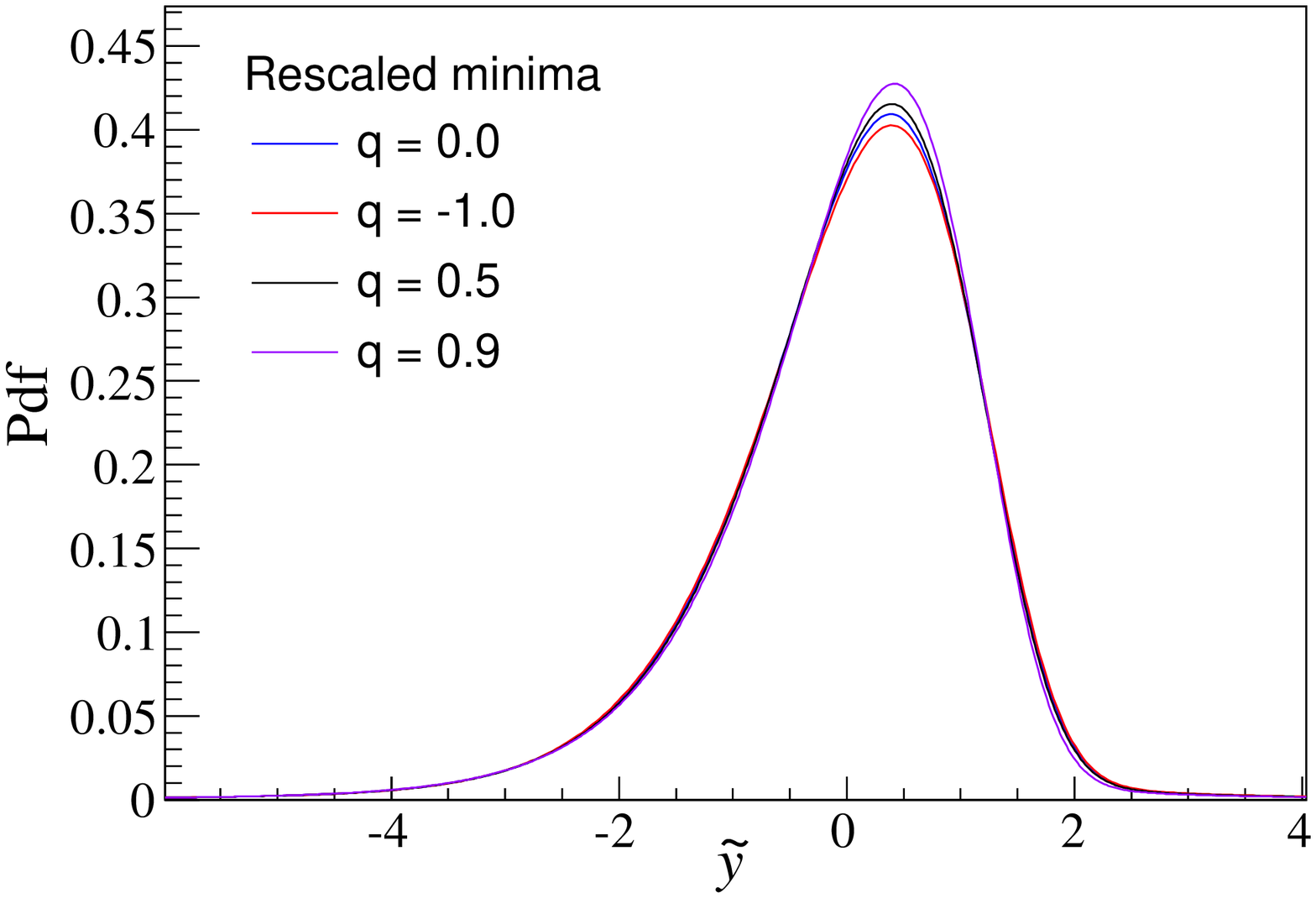}
\caption{Analytical prediction for minima distribution. Left: The mean is shifted to zero. Right: The distributions are rescaled to have zero mean and unit variance.}\label{minima-fig}
\end{figure}
Eq. (\ref{eq_etf}) predicts free energy fluctuations for $\beta < 1$. For $\beta\geq 1$, we shall assume the \textit{freezing scenario}. It appeared first as front velocity selection criteria in Fisher-Kolmogorov-Petrovsky-Piscounov type equations \cite{fisher1937wave,kolmo91kpp}. Later it emerged in the physics of disordered systems: first these defined on trees \cite{derrida1988polymers,krapivsky00kpp}, and then Euclidean-space ones \cite{chamon1996localization,castillo97dirac,carpentier2001glass}. Recently, it is proved by Madaule \textit{et. al.} for a large class of models \cite{madaule2013glassy}. To state it in our context, consider the random variable $y_\beta$ defined by $y_\beta := f - \beta^{-1}g$, where $g$ is a Gumbel variable \cite{gumbel2012statistics} independent of $f$. Equivalently,
\begin{equation}\label{ety-eq}
\overline{\exp(ty_\beta)} = \overline{\exp(tf)} \Gamma\left(1 + \frac{t}{\beta} \right) 
\end{equation}
By eq. \ref{eq_etf}, the above has a well-defined limit as $\beta \nearrow 1$. Now the freezing scenario claims it becomes $\beta$-independent for $\beta \geq 1$:
\begin{equation}\label{freez-eq}
\left.\overline{\exp(ty_{\beta})}\right\vert_{\beta \geq 1} \equiv  \left.\overline{\exp(ty_\beta)}\right\vert_{\beta \nearrow 1}.
\end{equation}
In particular, taking the $\beta\rightarrow\infty$ limit, we have $y_{\beta=1} = y_{\beta\rightarrow\infty} = f(\beta\rightarrow\infty) = y$, giving the minimum distribution. Its p.d.f., calculated by inverse-Fourier-transforming eq. (\ref{ety-eq}) is plotted in fig. (\ref{minima-fig}). Remark that $\mathbf{s}(t,\beta = 1, q)$ have poles only at $t = -2, -3, \dots$, (eq. (\ref{eq_s})), so the rightmost pole of $\overline{\exp(ty)}$ is at $t = -1$ and of order $2$, (eq. (\ref{ety-eq})). By inverse Fourier transform, this implies $\mathrm{Pdf}(y\rightarrow-\infty) \sim |y| e^{-|y|}$ for any $q$, confirming the universality of the Carpentier-Le Doussal tail \cite{carpentier2001glass}. 

We turn now to the\textit{ duality}. Authors of \cite{fyodorov2009statistical} observed for the FB model and the interval model that, if one continues \textit{analytically} $\overline{\exp(ty_\beta)}$ to $\beta \geq 1$, the unique result is \textit{self-dual}, \textit{i.e.}, invariant under $\beta\mapsto \beta^{-1}$. We stress that the self-dual solution is physically wrong for the phase $\beta \geq 1$, where the \textit{non-analytical} continuation eq. (\ref{freez-eq}) holds.
An important result of this work is that, $\overline{\exp(ty_\beta)}$ enjoys the same duality property, unaffected by finite domain effects. It follows from eq. (\ref{eq_etf}) and the self-duality of $\mathbf{s}(t,\beta,q)$ of eq. (\ref{eq_s}):
$$\mathbf{s}(t,\beta,q) = \mathbf{s}\left(t,\beta^{-1},q\right).$$
In fact, the term for $\lambda$ and for its transpose $\lambda' = \{(y,x): (x,y)\in\lambda\}$ are related by $\beta \leftrightarrow \beta^{-1}$. \footnote{The uniqueness of analytical continuation follows from the fact that $\mathbf{s}(t,\beta,q)$ is a sum of rational functions, \textit{convergent} when $\abs{q} < 1$ for all $t,\beta$ except for poles.} 
The co-presence of freezing and duality property supports the conjecture of \cite{fyodorov2009statistical} , which remains nonetheless intriguing from a general theoretical viewpoint. %   %  as it is known for the Liouville theory describing 2d Quantum Gravity , or in the minimal models describing 2d critical points 

\section{Numerical Study}\label{sect-num}
Numerical simulations of our models follow the same protocol as in \cite{fyodorov2009statistical}, sect. 6.1. The finite models defined at the beginning of sect. \ref{calc_sec} are simulated using Fast Fourier Transform. 
We will focus on: \textsf{(i)} the validity of analytical continuation leading to eq. (\ref{eq_etf}) and \textsf{(ii)} the freezing scenario by measuring the free energy variance $\overline{f^2}^c$ as a function of temperature.

We first study the free energy distribution in the $\beta<1$ phase, predicted by eq. (\ref{eq_etf}). To compare analytic and numerical values, the mean value should be shifted away from both. For numerical data, we measure the empirical mean value and subtract it from each sample: $f'_i := f_i - (\sum_{j=1}^n f_j) / n$ and calculate the Fourier transform $\sum_{i=1}^n \exp(\im f'_i t) / n$ of the sample distribution. On the analytical side, we calculate $ \overline{\exp(\im f t)} \exp(-\im \overline{f} t)$. The sum $\mathbf{s}(t,\beta,q)$ in eq. (\ref{eq_etf}) is calculated with the matrix product eq. (\ref{eq_s_mp}), and the first moment $\overline{f}$ using eqs. (\ref{eq_moments}) and (\ref{eq_f1_mp}).
\begin{figure}
\center
\includegraphics[scale=0.3]{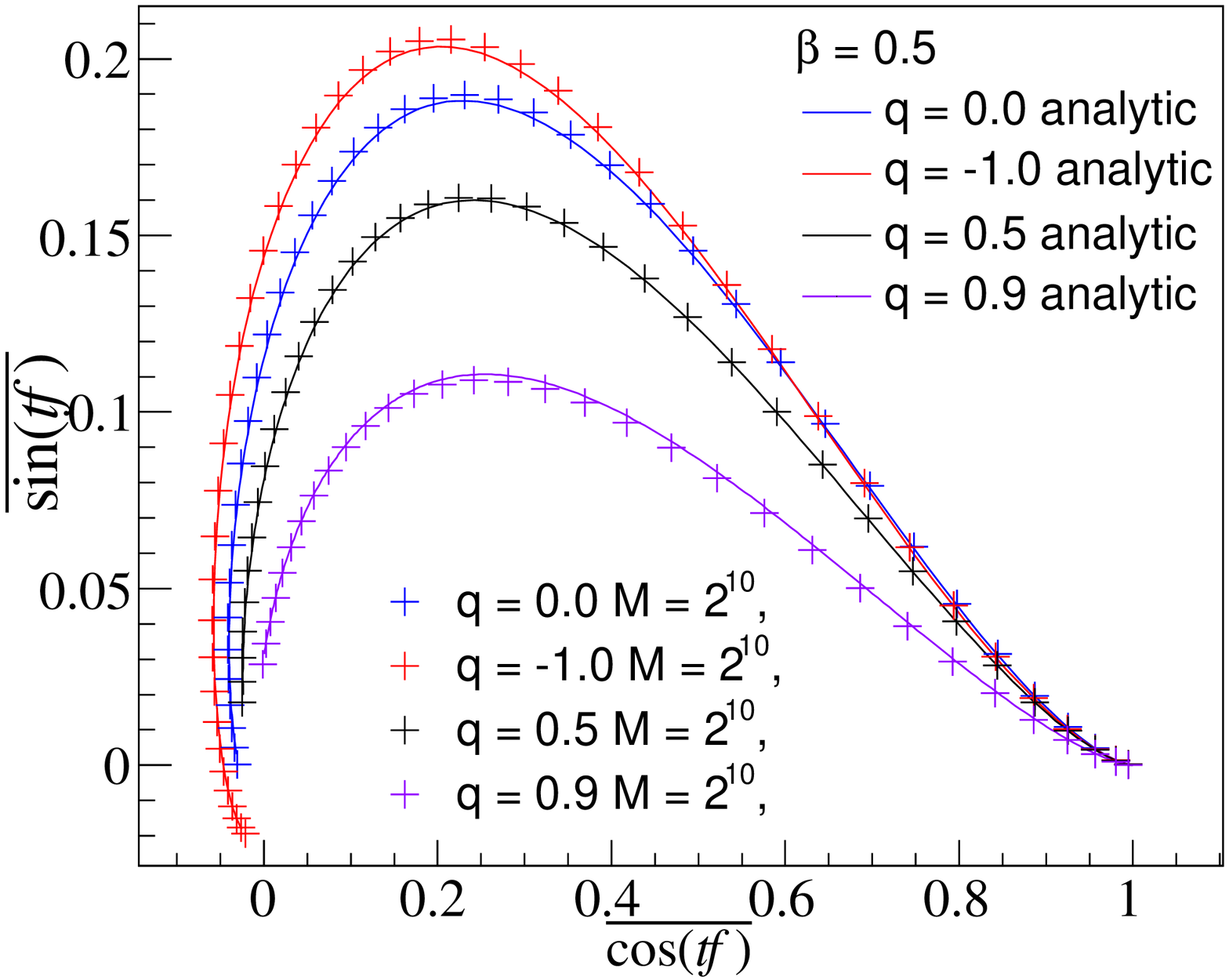}
\includegraphics[scale=0.3]{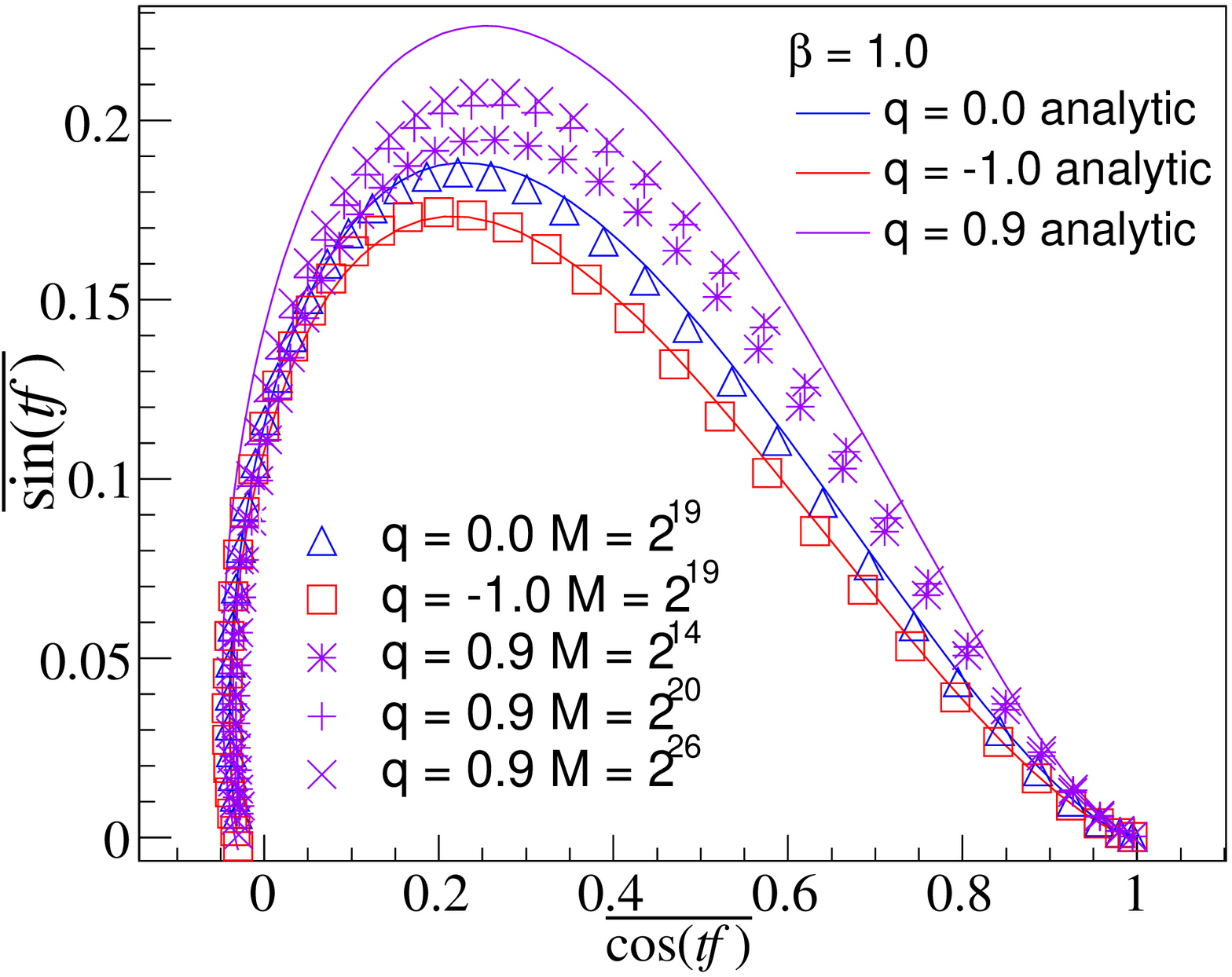}
\caption{Analytic vs. Numerical calculation of Fourier transform of free energy $\exp(\im t f)$, $t > 0$. Besides what is plotted, several other combinations $(q,\beta)$ are also checked. \textit{Left:} In the high temperature phase, here $\beta = 0.5$, the analytic-numerical agreement is excellent for all values of $q$ from $-1$ up to $0.9$. \textit{Right:} At the transition, finite size corrections amplify, especially for $q \nearrow 1$.} \label{hight_fig}
\end{figure}
In fig. (\ref{hight_fig}), we plot real and imaginary parts of the Fourier transform, so $t$ becomes an invisible parameter. This amounts to rescaling the distributions to have variance $1$, so that we concentrate here on the detailed shape of the distribution, \textit{e.g.}, its asymmetry.
The results validate the prediction of the free energy distribution in the high-$T$ phase.

\begin{figure}
\center
\includegraphics[scale=.6]{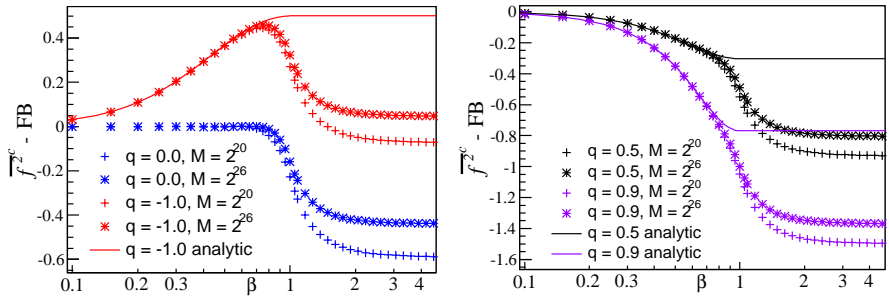}
\caption{The variance of $\overline{f^2}^c$ with the FB ($q = 0$) analytic values subtracted.  The analytic-numerics agreement is excellent in the high-$T$ phase and degrades when approaching the transition. In the low temperature phase, the plateau indicating freezing is observed for all numerical data. }
\label{gel-fig}
\end{figure}

Now we test the freezing scenario. We follow a strategy of \cite{fyodorov2009statistical} (sect. 6.4) and look at the variance of free energy $\overline{f^2}^c$ as a function of $\beta$. The $\beta<1$ part supplements the precedent numerics, while the $\beta > 1$ part tests the freezing scenario proper. The analytical prediction is calculated from the matrix product eqs. (\ref{eq_moments}), (\ref{eq_f1_mp}) and (\ref{eq_f2_mp}) for $\beta < 1$, and continued according to eq. (\ref{freez-eq}) to $\beta > 1$. The results are shown in fig. (\ref{gel-fig}). Although convergence in the $\beta>1$ phase is notoriously slow, we observe tantamount evidence of freezing for all values of $q$. 
The validity of the predicted free energy distribution in $\beta<1$ phase, plus the freezing scenario, entail the correctness of the prediction minima statistics. 

\section{Conclusion}
In this work, we mainly focused on the GFF on the unit disk with Dirichlet boundary condition. We proposed the minima distribution of the GFF on a centred circle of radius $\sqrt{q}$ for general $q\in(0,1)$ (fig. (\ref{minima-fig})). This was done by first proposing the exact free energy distribution of the Random Energy Model for $\beta<1$, tested with excellent agreement against numerics, fig. (\ref{hight_fig}), and then applying the freezing scenario, eq. (\ref{freez-eq}). Furthermore, as explained in sect. \ref{sec-freez}, the \textit{self-duality} enjoyed by our solution. Interpretation of results when $q\in[-1,0)$ and related problems on the sphere is also discussed (sect. \ref{intro}). The results presented here are the first that probe the effects of boundary conditions or finite volume. We show that these latter do affect the minimum distribution, but keep intact duality and freezing features. Their co-presence is observed for $\overline{\exp(ty_\beta)}$ of all solved 2d GFF models, as well as for velocity moments in decaying Burgers turbulence \cite{fyodorov2010freezing}. It is interesting to investigate its generality, origin and relation with the duality appearing in Random Matrix Theory \cite{dumitriua2002matrix} or Conformal Field Theory \cite{zamolodchikov1996conformal,dotsenko1985four}.

\textit{Acknowledgements.} We are grateful towards P. Le Doussal for a careful reading of the manuscript. We thank E. Bogomolny, A. De Luca, O. Giraud and C. Texier for useful discussions.

\appendix
\section{Matrix product form}\label{mp-append}
\begin{figure}
\includegraphics[scale=1]{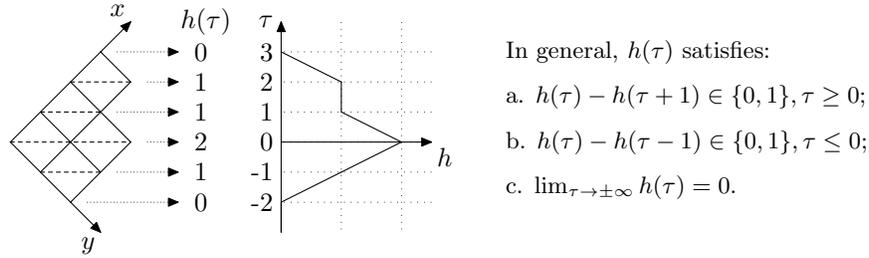}
\caption{\textit{Left}: An example illustrating the mapping between partitions $\lambda$ and paths $h(\tau)$. \textit{Right}: Conditions $h$ should satisfy to correspond to a partition.} \label{fig-russian}
\end{figure}
The infinite sum $\mathbf{s}(t,\beta,q)$ of eq. (\ref{eq_s}) is absolutely convergent for $|q| < 1$ and away from poles, but inefficient for practical use. Moreover, for $q = -1$, $\mathbf{s}$ is \textit{not} absolutely convergent, and the naive partial sums $\sum_{|\lambda| \leq N}$ oscillate $\gtrsim10\%$ even when $N \sim 80$. Fortunately, both problems can be resolved by a matrix product form of $\mathbf{s}(t,\beta,q)$, which we derive here.

First, we consider partitions as lattice walks $h(\tau), \tau\in\Z$, where $h(\tau)$ is the number of squares on the diagonal $x - y = \tau$ (see fig. (\ref{fig-russian}) for explanation). Observe that the product over boxes in eq. (\ref{eq_s}) telescopes along diagonals and becomes a product over $\tau$. We have indeed 
\begin{equation}
\mathbf{s}(t,\beta,q) := \sum_{\{h(\tau)\}} \prod_{\tau \in \Z} \frac{ q^{h(\tau)}(|\tau| \beta^{\mathrm{sgn}(\tau)} + t)}{|\tau|\beta^{\mathrm{sgn}(\tau)} +  \left(\beta + \beta^{-1}\right) h(\tau) + t} \label{eq_s_h}
\end{equation}
The sum is over all walks $h(\tau)$ satisfying conditions a, b and c of fig. (\ref{fig-russian}). The RHS resembles a path integral (or the partition function of a directed polymer), and can be written as a product of (transfer) matrices. To avoid infinite vector spaces, consider $\mathbf{s}^{(l)}$, the truncation of \ref{eq_s_h} to paths $h(\tau), \tau = -l, \dots, l$ such that $h(\pm l) = 0$ (\textit{i.e.}, we sum over all Young diagrams contained in the square $\{ x < l, y < l \}$). Using the auxiliary space spanned by $ \vert h \rangle, h = 0,1,2 \dots l$, we have
\begin{eqnarray}
\mathbf{s}^{(l)} =  \langle 0 \vert U D_{l-1} U\dots D_{1} U \vert C \vert
   L E_1 \dots L E_{l-1} L  \vert  0  \rangle  \label{eq_s_mp} \\  
  U = {I} + \sum_{h = 0}^{l-1} \vert h \rangle\langle h + 1 \vert,  D_j(\beta) = \sum_{n = 0}^{l} \frac{q^h \vert h \rangle\langle h \vert}{1 + h(\beta + \beta^{-1})(j\beta + t)^{-1}}, \\
   L = U^\dagger, E_j = D_j^\dagger\left(\beta^{-1}\right), C = D_0 = E_0. 
\end{eqnarray}
Remark that $D_j,E_j$ and $C$ are diagonal (they generate factors in eq. (\ref{eq_s_h}), $j = |\tau|$), while $U$ and $L$ are nearly so (they implement conditions a and b in fig. (\ref{fig-russian}). This enables the sum $\mathbf{s}(l)$ over $O(4^l)$ partitions to be calculated in $O(l^2)$ time and $O(l)$ space, so as to achieve in practice the convergence $\mathbf{s}^{(l)}\rightarrow\mathbf{s}$, $l\nearrow \infty$. Using $l \sim 10^3$, we observed 4 decimal precisions, even for $q = -1$. Below, $\mathbf{s}$ will be understood as $\mathbf{s}^{(l)}$ for $l$ sufficiently big. 

To calculate moments of $f$, we need $t$-derivatives of eq. (\ref{eq_s_mp}) at $t = 0$. In fact, from eq. (\ref{eq_etf}) we have
\begin{equation}
s_k := \left.\left( \partial_t^k \ln \mathbf{s}\right)\right|_{t=0}\; \Rightarrow \;\overline{f^k}^c = s_k + \beta^k \left(\ln\Gamma(x)\right)^{(k)}_{x=1}, \; \beta \leq 1. \label{eq_moments}
\end{equation}
For example, $\overline{f}^c = s_1 - \beta \gamma_E, \overline{f^2}^c = s_2 +  \beta^2 \pi^2/6$, and so on. 
Differentiating $\mathbf{s}$ is done by Leibniz rule, and the result is written again as matrix products. For the first moment, one has 
\begin{equation} 
s_1 = \langle 0 \vert U D_{l-1}^{(0)} U\dots D_{1}^{(0)} U \vert C^{(1)} \vert
   L  E_{1}^{(0)} \dots L  E_{l-1}^{(0)} L  \vert  0  \rangle  \label{eq_f1_mp}
\end{equation}
where $D_j^{(0)} = D_j\vert_{t=0}$, similarly for $E_{j}^{(0)}$, and $C^{(1)} =\left.\left( \partial_t C\right)\right\vert_{t=0}$.

For the $k$-th moment, one need an auxiliary space of $(l+1)k$-dimension (to bookkeep the positions hit by derivatives), spanned by $\vert h,d\rangle := \vert h\rangle \otimes \vert d \rangle, h = 0,\dots,l, d = 0,\dots,k-1$. The general expression is cumbersome, and we will give that of $k=2$: 
\begin{eqnarray}
s_2 + s_1^2= 
 \langle 0,0 \vert U D_{l-1}^{(1)} U\dots D_{1}^{(1)} U \vert C^{(2)} \vert
   L  E_{1}^{(1)} \dots L  E_{l-1}^{(1)} L  \vert 0 ,0 \rangle  \label{eq_f2_mp} \\
  D^{(1)}_j = D_j^{(0)} \otimes \mathbb{I} + \partial_t D_j \vert_{t = 0} \otimes \vert 0  \rangle\langle 1\vert, E = D^{\dagger}(\beta \rightarrow \beta^{-1}) \\
  C^{(2)} = 2 \left.\left( \partial_t C\right)\right\vert_{t=0}
   \otimes (\vert 0 \rangle\langle 1\vert  +  \vert 1 \rangle\langle 0\vert)
   +  \left.\left( \partial_t^2 C\right)\right\vert_{t=0} \otimes \vert 0 \rangle\langle 0\vert
\end{eqnarray}

%To end this section, we discuss briefly convergence issues. %Furthermore, for any $b > -\min(\beta,\beta^{-1})$, on the half plane $\{t \vert \Re(t) \geq b\}$, $\mathbf{s}$ can be uniformly bounded. We checked empirically positivity and convexity of (\ref{eq_etf}, \ref{ety-eq}). 
%The more tricky case is 

\section{High temperature expansion}\label{sec_expan}
As an another check of eq. (\ref{eq_etf}), we compute the high temperature expansion of the free energy variance $\overline{f^2}^c$ at order $\beta^2$. As is shown in \cite{fyodorov2009statistical}, (eq. (C.5)), 
$\overline{f^2}^c = \beta^2 \sum_{k\neq0} \frac{\mu_k^2}{2} + O(\beta^4)$. We have $\mu_k = (1 - q^{|k|}) / k$ (eq. \ref{eq-modes}), which implies  
\begin{eqnarray}\label{eq_highT}
\overline{f^2}^c = \left( \mathrm{Li}_2(q^2)-2\mathrm{Li}_2(q) + \frac{\pi^2}{6} \right) \beta^2 + O(\beta^4),  \mathrm{Li}_m(x) := \sum_{k > 0} \frac{x^k}{k^m}.
\end{eqnarray}
Now we derive this from eq. (\ref{eq_etf}). The $\Gamma(1+t\beta)$ part gives the contribution $\frac{\pi^2}{6}$, The sum eq. (\ref{eq_s}) is developed at $t = \beta = 0$ as: 
$\sum_\lambda q^{|\lambda|}\prod_{(x,y)\in\lambda} \frac{x\beta^2 + y + t\beta}{(x+1)\beta^2 + (y+1) + t\beta } = 1 + t(m_1 \beta + \dots)  +  t^2(m_2 \beta^2 + \dots) + \dots,$
where $\dots$ are higher order terms.

Let us begin by $m_1$. For any non-empty partition, the box $(x=0,y=0)$ has already nominator $t\beta$; so the other boxes should all be evaluated at $t = \beta = 0$ when contributing to $m_1$. So, only $\lambda$'s with one column contribute; otherwise the box $(x=1,y=0)$ would have a nominator $(\beta^2 + t\beta)$ that vanishes. Therefore
\begin{equation}\label{m1_eq}
m_1 = \sum_{k>0} q^k \prod_{y = 0}^{k-1} \frac{y}{y+1} = \sum_{k>0} \frac{q^k}{k} = \mathrm{Li}_1(q) 
\end{equation}
Similarly only one-column and two-column partitions contribute to $m_2$. A two-column partition contributes by the product of all boxes with $y > 0$ evaluated at $\beta = t = 0$, yielding
\begin{eqnarray}
m_{2,2} = \sum_{k \geq l > 0} q^{k + l}  
\prod_{y = 0}^{k-1} \frac{y}{y+1}\prod_{y = 0}^{l-1} \frac{y}{y+1} = \frac{1}{2}\mathrm{Li}_1(q)^2 + \frac{1}{2}\mathrm{Li}_2(q^2) \label{m22_eq}
\end{eqnarray}
Finally, contributions from each one column partition is the sum over where the second $t\beta$ term comes from: 
\begin{eqnarray}
m_{2,1} = \sum_{k > 0} \frac{q^{k}}{k} \left[-1 + \sum_{y=1}^{k-1} \left(\frac{1}{y} - 
\frac{1}{y + 1}  \right) \right] = - \mathrm{Li}_2(q) \label{m21_eq}
\end{eqnarray}
Combining eqs. \ref{m1_eq} to \ref{m21_eq} one has $2m_2 - m_1^2 = 2(m_{2,1} + m_{2,2}) - m_1^2 =  \mathrm{Li}_2(q^2)-2\mathrm{Li}_2(q)$ as desired. 
\bibliography{rems}
\end{document}